\documentclass[twocolumn,showpacs,preprintnumbers,amsmath,amssymb,aps,prd,superscriptaddress]{revtex4-1}
\makeatletter

\newcommand{\Rmnum}[1]{\expandafter\@slowromancap\romannumeral #1@}
\makeatother
\usepackage[colorlinks,linkcolor=blue,anchorcolor=blue,citecolor=blue,urlcolor=blue,breaklinks=true]{hyperref}

\usepackage{epsfig}
\usepackage{natbib}
\usepackage{epstopdf}
\usepackage{mathrsfs}
\usepackage{slashed}
\usepackage{amsmath}
\usepackage{verbatim}
\usepackage{graphicx}
\usepackage{amssymb}
\usepackage{amsmath}
\usepackage{psfrag}
\usepackage{array}
\usepackage{lipsum}
\usepackage{float}
\usepackage[dvipsnames]{xcolor}

\usepackage{bigstrut,multirow,rotating}
\usepackage{booktabs}
\usepackage[normalem]{ulem}

\begin{document}

\title{Jet momentum reconstruction in the QGP background with machine learning}
\author{Ran Li}
\affiliation{Institute of Frontier and Interdisciplinary Science, Shandong University, Qingdao, 266237, China}
\author{Yi-Lun Du}
\email{yilun.du@iat.cn}
\affiliation{Shandong Institute of Advanced Technology, Jinan 250100, China}
\author{Shanshan Cao}
\email{shanshan.cao@sdu.edu.cn}
\affiliation{Institute of Frontier and Interdisciplinary Science, Shandong University, Qingdao, 266237, China}
\begin{abstract}

We apply a Dense Neural Network (DNN) approach to reconstruct jet momentum within a quark-gluon plasma (QGP) background, using simulated data from PYTHIA and Linear Boltzmann Transport (LBT) Models for comparative analysis. We find that medium response particles from the LBT simulation, scattered out of the QGP background but belonging to medium-modified jets, lead to oversubtraction of the background if the DNN model is trained on vacuum jets from PYTHIA simulation. By training the DNN model on quenched jets generated using LBT or the combination of jet samples from PYTHIA and LBT, we significantly reduce this prediction bias and achieve more accurate background subtraction compared to conventional Area-based and Constituent Subtraction methods widely adopted in experimental measurements. We further study the performance of these machine learning models on evaluating the nuclear modification factor of jets, and find that while the unfolding procedure is necessary for correcting residuals in reconstructed jet momenta, models trained on samples incorporating quenched jets still achieve superior accuracy than those trained on vacuum jets even after unfolding. 
\end{abstract}

\maketitle


\section{Introduction} 

Jets, collimated sprays of particles originating from splittings and hadronization of energetic partons produced in hard collisions between nucleons, serves as an excellent probe of the quark-gluon plasma (QGP) created in relativistic heavy-ion collisions~\cite{Gyulassy:2004zy,Jacobs:2004qv,Busza:2018rrf,Elfner:2022iae}. Scatterings between jet partons and the QGP can result in significant suppression, or ``quenching", of the jet spectra in nucleus-nucleus (A+A) collisions relative to their baselines in proton-proton ($p+p$) collisions~\cite{Wang:1992qdg}. The jet transport coefficient inside the QGP extracted from the jet quenching data appears an order of magnitude larger than that inside a cold nucleus probed by deeply inelastic scatterings~\cite{JET:2013cls,JETSCAPE:2021ehl,Xie:2022ght,Chen:2024epd}, revealing the partonic degrees of freedom inside the QGP. Considerable efforts have been dedicated into understanding the interaction dynamics between jets and the QGP~\cite{Bass:2008rv,Wiedemann:2009sh,dEnterria:2009xfs,Armesto:2011ht,Qin:2015srf,Majumder:2010qh,Blaizot:2015lma,Mehtar-Tani:2013pia,Cao:2020wlm,Cao:2022odi,Apolinario:2022vzg} and developing sophisticated Monte-Carlo event generators to simulate jet scatterings through the QGP~\cite{Cao:2024pxc,Schenke:2009gb,Zapp:2013vla,Casalderrey-Solana:2014bpa,Cao:2017qpx,JETSCAPE:2017eso,Putschke:2019yrg,Luo:2023nsi,Karpenko:2024fgg}. These enable us to extend studies on jets from the quenching of their yields to nuclear modification of their substructures~\cite{Apolinario:2024equ,Casalderrey-Solana:2016jvj,Tachibana:2017syd,KunnawalkamElayavalli:2017hxo,Milhano:2017nzm,Chen:2020tbl,Park:2018acg,Luo:2018pto,Casalderrey-Solana:2019ubu,Chang:2019sae,Tachibana:2020mtb,Yang:2023dwc,Xing:2024yrb}, and from improving model descriptions of the jet data to quantitatively extracting various QGP properties using the jet data~\cite{Liu:2023rfi,Karmakar:2023ity,Karmakar:2024jak}. 

While it is straightforward to track jet particles from their initial production to their subsequent splittings and hadronization in theoretical simulations, it is a great challenge to identify jet particles in relativistic nuclear collision experiments due to the background of particles produced from multi-parton interactions in the initial state or the QGP medium in the final state. Proper ways of subtracting these backgrounds in experiments are essential for correctly extracting jet observables for a meaningful comparison to theoretical predictions. Therefore, tremendous dedication has been devoted in designing efficient background subtraction methods. For instance, for jet momentum reconstruction, conventional schemes such as the Area-based method~\cite{Cacciari:2011ma,Abelev:2013kqa} and the Constituent Subtraction method~\cite{Berta:2019hnj,JetToyHI} have been developed. They rely on estimating and subtracting the average background contributions after excluding the leading jets, however, tend to discard subtle features of individual jets, and thus often result in large residual fluctuations. While being effective in $p+p$ collisions, these methods may struggle in heavy-ion collisions due to the complicated interactions between jets and the QGP. In the presence of the QGP, subtraction can be biased by recoil particles from the medium and other noise sources, leading to inaccurate reconstruction of jet momentum and substructures, especially for jets with low transverse momenta ($p_\mathrm{T}$). This highlights the necessity for establishing more flexible and accurate approaches that can adapt to complex backgrounds, and capture rich features of individual quenched jets in heavy-ion collisions.

Recent advancements in machine learning (ML) have shown significant promise in heavy-ion physics~\cite{PhysRevLett.114.202301,PhysRevC.94.024907,bernhard2019bayesian,PhysRevC.101.024911,PhysRevLett.126.202301,PhysRevC.103.054909,PhysRevC.97.014907,Liu:2023rfi,bhalerao2015principal,mazeliauskas2015subleading,pang2018equation,liu2019principal,du2020identifying,PhysRevC.104.044902,mallick2021estimation,wang2021finding,Boehnlein:2021eym,xiang2022determination,shi2022heavy,li2023deep,soloveva2024extraction,zhou2024exploring}, particularly in the study of jet quenching in QGP~\cite{Du:2024gN}. 
Specifically, ML techniques have been successfully applied to the classification between quark and gluon jets~\cite{du2021classification,Chien:2018dfn}, identification of quenched jets~\cite{apolinario2021deep,Liu:2022hzd,lai2021information,romao2023jet,qureshi2024model}, prediction of jet energy loss~\cite{Du:2020pmp,du2021jet,Li:2025tqr} and locating jet creation vertices~\cite{yang2023deep}, demonstrating their potential to capture subtle features of jet modifications in a QGP environment. Regarding to background subtraction, ML also offers a powerful approach to extract the momenta of jets embedded in QGP backgrounds by learning patterns of jet signals inside backgrounds on a jet-by-jet basis~\cite{haake2019machine,acharya2024measurement,mengel2023interpretable}. In Ref.~\cite{haake2019machine}, jets from PYTHIA simulation~\cite{Sjostrand:2006za,Sjostrand:2014zea} are placed in a thermal background for model training, suggesting that the ML-based reconstruction approach achieves superior precision in jet momentum reconstruction compared to the conventional Area-based method. This ML-based method has been further applied to the ALICE experimental data where the nuclear modification factor ($R_\mathrm{AA}$) has been obtained for jets with $p_\mathrm{T}$ down to 20 to 40~GeV~\cite{acharya2024measurement}. Considering that PYTHIA simulates jet production in $p+p$ collisions, different medium modification effects in A+A collisions, such as the variation of the quark-to-gluon-jet fraction, medium-induced gluon emission, and jet-induced medium excitation, are incorporated separately in modifying the training data, based on which the systematic error of $R_\mathrm{AA}$ from implementing different ML models is estimated in Ref.~\cite{acharya2024measurement}. 

For an improved treatment of medium modification of jets, in this work, we will adopt the state-of-the-art Linear Boltzmann Transport (LBT) model~\cite{Luo:2023nsi} to generate data of quenched jets for training neural networks, and compare its performance to that of both conventional methods and ML models trained with the PYTHIA data. We will not only demonstrate the better precision of our current ML-based method than pioneer ones in predicting the jet momentum and $R_\mathrm{AA}$, but also reveal the physics origin of this superior performance. We will show that a ML model trained using the PYTHIA data alone is incapable of identifying recoil particles produced from jet-medium interactions, and thus inevitably overestimates the QGP background and underestimates the jet momentum and $R_\mathrm{AA}$. This bias cannot be corrected by the unfolding procedure. Therefore, it's necessary to use event generators with realistic jet quenching effects in training ML models for jet analysis in future heavy-ion experiments.

\section{Simulation Models}
\label{sec: sim models}
We train neural networks using simulation models that capture the essential features of jet formation and their interactions with the QGP. Below, we briefly introduce the three key models employed in this work: the PYTHIA model of jet production in $p+p$ collisions, the LBT model for jet-QGP interactions, and a toy thermal model for mimicking the QGP background.

PYTHIA~\cite{Sjostrand:2006za,Sjostrand:2014zea} is a widely-used event generator for simulating high-energy particle collisions, such as those conducted at the Large Hadron Collider (LHC). It can model the formation of jets in quantum chromodynamics (QCD) processes that start with hard scatterings between partons from projectile and target nucleons. Highly virtual partons produced from these hard scatterings then evolve to lower scales by sequential splittings. This is known as parton shower. As the daughter partons from these splittings approach the hadronization scale, they are converted into hadrons based on the Lund string model. One may further apply jet finding algorithms, such as the anti-$k_\mathrm{T}$~\cite{Cacciari:2008gp} and Cambridge–Aachen~\cite{Dokshitzer:1997in,Wobisch:1998wt} algorithms to cluster these hadrons into jets and compare their observables to experimental data. In the context of studying nuclear modification of jets in heavy-ion collisions, PYTHIA provides a baseline of $p+p$ collisions that excludes jet quenching effects. 

The Linear Boltzmann Transport (LBT) model~\cite{cao2016linearized,Luo:2023nsi} simulates elastic and inelastic scatterings between jet partons and thermal partons inside the QGP based on the Boltzmann equation. The elastic scattering rates of jet partons are calculated using the leading-order matrix elements of $2\rightarrow 2$ processes, with all possible partonic scattering channels taken into account~\cite{He:2015pra}; and the inelastic scattering rates are related to the spectra of medium-induced gluon emissions from jet partons, which can be evaluated using the higher-twist energy loss formalism~\cite{Guo:2000nz,Wang:2001ifa,Zhang:2003yn,Majumder:2009ge,Zhang:2003wk,Zhang:2004qm,du2018revisiting,Zhang:2019toi,Sirimanna:2021sqx}. The LBT model keeps track of not only the initially produced jet partons and their emitted gluons inside the QGP, but also the medium partons that are scattered out of the thermal background (named as ``recoil" partons) and the energy holes left behind (named as ``negative" partons). These recoil and negative partons constitute medium response to jet propagation in LBT, which is naturally included in medium-modified jets and has been shown to be essential to jet observables related to soft particles at large angles relative to jet axes. The initially produced jet partons, their emitted gluons, recoil partons and negative partons all belong to ``jet partons" produced by the LBT model. Contributions from negative partons are subtracted from those of other jet partons for all jet observables. By combining the LBT model with a QGP background simulated by the (3+1)-dimensional viscous hydrodynamic model CLVisc~\cite{Pang:2018zzo,Wu:2021fjf}, one can achieve good descriptions of the quenching of hadrons and jets~\cite{Xing:2019xae,He:2018xjv}, their collective flows~\cite{He:2022evt}, and jet substructures~\cite{Luo:2018pto}. In the present work, we use the LBT model to further evolve partons produced by PYTHIA inside the QGP. Its output data allow ML models to learn the complex energy dissipation patterns and structural changes in jets due to their interactions with the QGP. Consequently, the LBT model provides a valuable tool for developing and validating ML approaches that aim to accurately reconstruct jet observables in relativistic heavy-ion experiments. 

To study the ML techniques in reconstructing jet momentum in the presence of a QGP background, we embed jet partons produced by either PYTHIA or PYTHIA+LBT into a background generated by a thermal toy model~\cite{JetToyHI}. For each jet event, the thermal model is tuned to produce  $\pi^\pm$ particles whose momenta follow a Boltzmann distribution with a total multiplicity of 1538 and an average transverse momentum $\langle p_\mathrm{T} \rangle = 0.5696$~GeV per unit of rapidity within the range of $|y| < 1$, mimicking the realistic soft charged hadron environment created in central (0-10\%) Pb+Pb collisions at $\sqrt{s_\mathrm{NN}} = 5.02$~TeV~\cite{acharya2020production}. Note that the power-law tail of particle distributions observed in experiments originates from jets, and the combination of thermal and jet particles should naturally produce the full momentum distribution with the power-law tail. In principle, one can also sample background particles from the freezeout hypersurface of the same hydrodynamic medium used for simulating jet-QGP interactions. However, for the purpose of comparing to earlier studies on this topic~\cite{haake2019machine,acharya2024measurement}, we keep using the thermal model to generate the background, which should also be sufficient for a proof-of-principle study on applying ML techniques to jet reconstruction.

In this work, we start with 360k PYTHIA~8 events with the Monash 2013 tune~\cite{Skands:2014pea}. The momentum exchange triggers of hard collisions are uniformly distributed within the range of [20, 200]~GeV. Final state partons from these PYTHIA events are then evolved using the LBT model coupled to a QGP medium for central (0-10\%) Pb+Pb collisions at $\sqrt{s_\mathrm{NN}} = 5.02$~TeV. The vertices of hard collisions in heavy-ion collisions are sampled using the Monte-Carlo Glauber model. Each parton is assumed to stream freely before its formation time and the thermalization time of the QGP (0.6~fm). The formation time of a parton is evaluated by summing over the splitting times of its ancestors in PYTHIA vacuum showers~\cite{Zhang:2022ctd}. After that, these partons interact with the QGP within LBT until the local temperature of the medium drops below the hadronization temperature (165~MeV). The strong coupling constant in LBT is set as $\alpha_\mathrm{s}=0.15$. Jet reconstruction is performed using the anti-$k_\mathrm{T}$ algorithm with a radius parameter of $R = 0.4$ for both PYTHIA and PYTHIA+LBT jets, either with or without the thermal background contributions. For analysis, we select jets within the transverse momentum range of $5~\mathrm{GeV} < p_\mathrm{T} < 160~\mathrm{GeV}$ and pseudorapidity range of $|\eta| < 1 - R$. For the rest of this work, we will use the notation ``LBT" to abbreviate ``PYTHIA+LBT" in describing how we generate the jet data. On the other hand, the notation ``PYTHIA+LBT" will be saved for describing the combination of jet data generated by PYTHIA simulation and (PYTHIA+)LBT simulation in training ML models.

\section{Background Subtraction Methods}
\label{sec: bkg methods}

\subsection{Area-based Method}
Two conventional background subtraction methods and a ML method will be used for a comparative study. The Area-based method~\cite{Cacciari:2011ma,Abelev:2013kqa} is a widely adopted approach for jet momentum reconstruction, which is effective in mitigating impact of the background noise, such as contributions from underlying events in high-energy collisions. It first estimates the density of the background as the median value of transverse momentum densities of clusters reconstructed using the $k_\mathrm{T}$ algorithm~\cite{Catani:1993hr,Ellis:1993tq}, with a few (two in this work) hardest clusters excluded:
\begin{equation}
\rho= \mathrm{median} \left({p_{\mathrm{T}, i}}/{A_i}\right), 
\end{equation} 
where $p_{\mathrm{T}, i}$ is the $p_\mathrm{T}$ of the $i$-th cluster, and $A_i$ is its area given by the ghost particle method. With this background density, the $p_\mathrm{T}$ of an anti-$k_\mathrm{T}$ jet is then given by
\begin{equation}
p^{\mathrm{Area-based}}_{\mathrm{T,\, jet}}= p^{\mathrm{within\, bkg}}_{\mathrm{T,\, jet}} - \rho  A,
\end{equation} 
where the first part on the right denotes the $p_\mathrm{T}$ of a jet reconstructed using the anti-$k_\mathrm{T}$ algorithm in the presence of the background, and $A$ denotes the area of this jet. 

\subsection{Constituent Subtraction Method}
\label{sec:const}

Unlike the Area-based method which subtracts a smooth background from a jet, the Constituent Subtraction method~\cite{Berta:2019hnj,JetToyHI} is designed to reduce the background noise by adjusting momenta of individual particles or removing individual particles within an event. This method is particularly useful in high-background environments, such as heavy-ion collisions, where the large number of soft background particles can introduce clustering biases~\cite{cacciari2008catchment}. To reduce these biases, the Constituent Subtraction method subtracts the background at the particle level before performing jet clustering, thereby minimizing its impact on finding ``correct" jets.

The Constituent Subtraction method starts with estimating the average background density $\rho$ of an event in the same way as discussed earlier for the Area-based method. Then, to emulate the background particle distribution, a large number of soft ghost particles are introduced, which uniformly cover the rapidity-azimuthal-angle $(y, \phi)$ plane under investigation. Each ghost particle here is assigned a transverse momentum of $p_\mathrm{T}^g = \rho A^g $, where $A^g=0.001$ is the area of a ghost particle. The total transverse momentum of ghost particles approximates that of the background.

For each pair of (real) particle $i$ and ghost $k$, their distance is defined as  
\begin{equation}
\centering
  \Delta R_{i,k}  = p^{\alpha}_{\mathrm{T},i}  \sqrt{(y_{i}-y_{k}^{g})^2+(\phi_{i}-\phi_{k}^{g})^2},   
\end{equation}
where $\alpha $ is a free parameter, set to 1 here. This distance incorporates a weight of the particle $p_\mathrm{T}$, considering that background particles generally have lower $p_\mathrm{T}$ than particles initiated by hard scatterings. By sorting these distances in an ascending order, the momentum removal begins with the smallest distance. For each pair, the transverse momenta of particle $i$ and ghost $k$ are adjusted as follows:
$$
\begin{array}{l}
p_{\mathrm{T}, i} \to p_{\mathrm{T}, i} - p_{\mathrm{T}, k}^{g}, \quad p_{\mathrm{T}, k}^{g} \to 0, \quad \text{if} \quad p_{\mathrm{T}, i} \geq p_{\mathrm{T}, k}^{g};\\
p_{\mathrm{T}, k}^{g} \to p_{\mathrm{T}, k}^{g} - p_{\mathrm{T}, i},   \quad  p_{\mathrm{T}, i} \to 0,\quad \text{otherwise.} 
\end{array}
$$ 
The removal process stops when $\Delta R_{i,k} > \Delta R^{\mathrm{max}} $ where $\Delta R^{\mathrm{max}}$ is a free parameter controlling the maximum range of neighboring ghosts considered for subtraction. We set  $\Delta R^{\mathrm{max}}$ as $\infty$ in this work. After these momentum subtractions, jets are reconstructed from particles with nonzero momenta using the anti-$k_\mathrm{T}$ algorithm, allowing for a straightforward determination of their $p_\mathrm{T}$.

By locally adjusting the momentum of each particle based on the background density, this Constituent Subtraction method is expected to isolate the jet signal while preserving the overall jet momentum and substructure. This feature is crucial for analyzing jet properties in high-background environments.

\subsection{Machine Learning Method}
\label{sec: ML method}

Dense neural networks (DNNs) are powerful tools for mapping input observations to output targets. Unlike conventional methods, neural networks do not require explicit mathematical models or intricate pre-computations. Instead, they autonomously extract features and establish mapping rules by learning from large datasets. This makes them particularly well-suited for problems involving high-dimensional and complex data structures.

In this work, we leverage neural networks to predict the jet momentum, following the approach in Ref.~\cite{haake2019machine} for the purpose of straightforward comparison. The input jet observables used in the model include:
\begin{itemize} 
\item Uncorrected jet momentum reconstructed by the jet finding algorithm;
\item Jet transverse momentum, corrected by the Area-based method;
\item Jet substructure observables: mass, angularity, momentum dispersion, and the difference between the momenta of the leading and sub-leading constituent tracks (LeSub); 
\item Number of constituents within the jet;
\item Mean and median transverse momenta of all constituents; 
\item Transverse momenta of the first ten hardest particles within the jet.
\end{itemize}
The target value of the jet momentum after the background subtraction, $p^{\mathrm{target}}_{\mathrm{T,\, jet}}$, is defined as the $p_\mathrm{T}$ sum of the raw jet particles generated by the PYTHIA or LBT simulation within the jet reconstructed in the presence of the QGP background: 
\begin{equation}
\label{eq: pT}
\centering
   p^{\mathrm{target}}_{\mathrm{T,\, jet}} =  \sum\nolimits_{i\in \mathrm{PYTHIA/LBT}} p_{\mathrm{T},i},
\end{equation}
where $i$ labels constituents within the reconstructed jet. Therefore, this $p^{\mathrm{target}}_{\mathrm{T,\, jet}}$ represents the $p_\mathrm{T}$ contributed by the ``true" jet partons originating from the PYTHIA or LBT model, excluding contributions from the background particles. We emphasize that in the presence of medium response, all particles generated by the LBT model are regarded as jet constituents. This includes particles produced by PYTHIA vacuum showers and placed into LBT, as well as medium-induced gluons, recoil particles, and negative particles generated by LBT. Contributions from negative partons are subtracted from those of other jet partons for calculating the jet $p_\text{T}$.

The DNN architecture employed in this study consists of three hidden layers containing 100, 100, and 50 nodes, respectively. Each neuron uses the Rectified Linear Unit (ReLU) activation function, defined as ReLU($x$) = max($x$, 0), to introduce non-linearity and efficiently model complex relationships in the data. The loss function chosen for training is the LogCosh function, log[cosh($x$)], which behaves as $x^2$/2 for small $x$ and $\mathrm{abs}(x) - \mathrm{log}(2)$ for large $x$. This property makes it robust to outliers while maintaining smooth sensitivity to small residuals. The Adam optimization algorithm~\cite{kingma2014adam} is employed to minimize the loss function during training, offering adaptive learning rates for efficient convergence. The dataset is split into training and testing subsets, with 80\% of the simulated jet data used for training the DNN model and the remaining 20\% reserved for testing its performance. This division ensures the model's ability to generalize its predictions to unseen data while preventing overfitting.

\section{Jet Momentum Reconstruction}
\label{sec:momentum}

\begin{figure}[tbp!]
\centering
\includegraphics[width=0.48\textwidth]{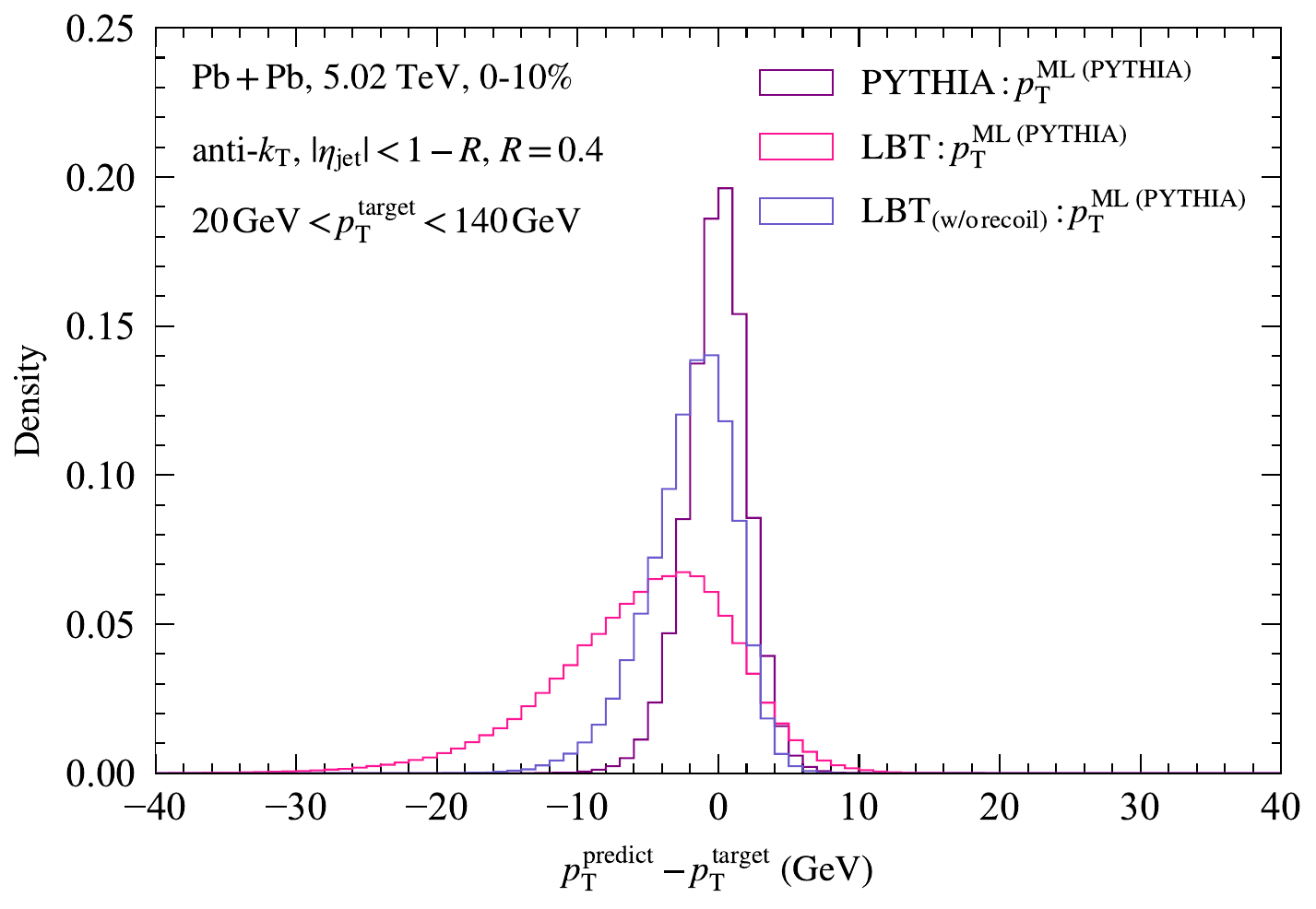}
\caption{(Color online) Residual distributions of the jet $p_\mathrm{T}$ predicted by the ML model trained on the PYTHIA data, compared between testing on datasets of PYTHIA, LBT, and LBT without recoil (and negative) particles, respectively.}
\label{fig: pT_error_pythia_training} 
\end{figure}

\begin{figure}[tbp!]
\centering
\includegraphics[width=0.48\textwidth]{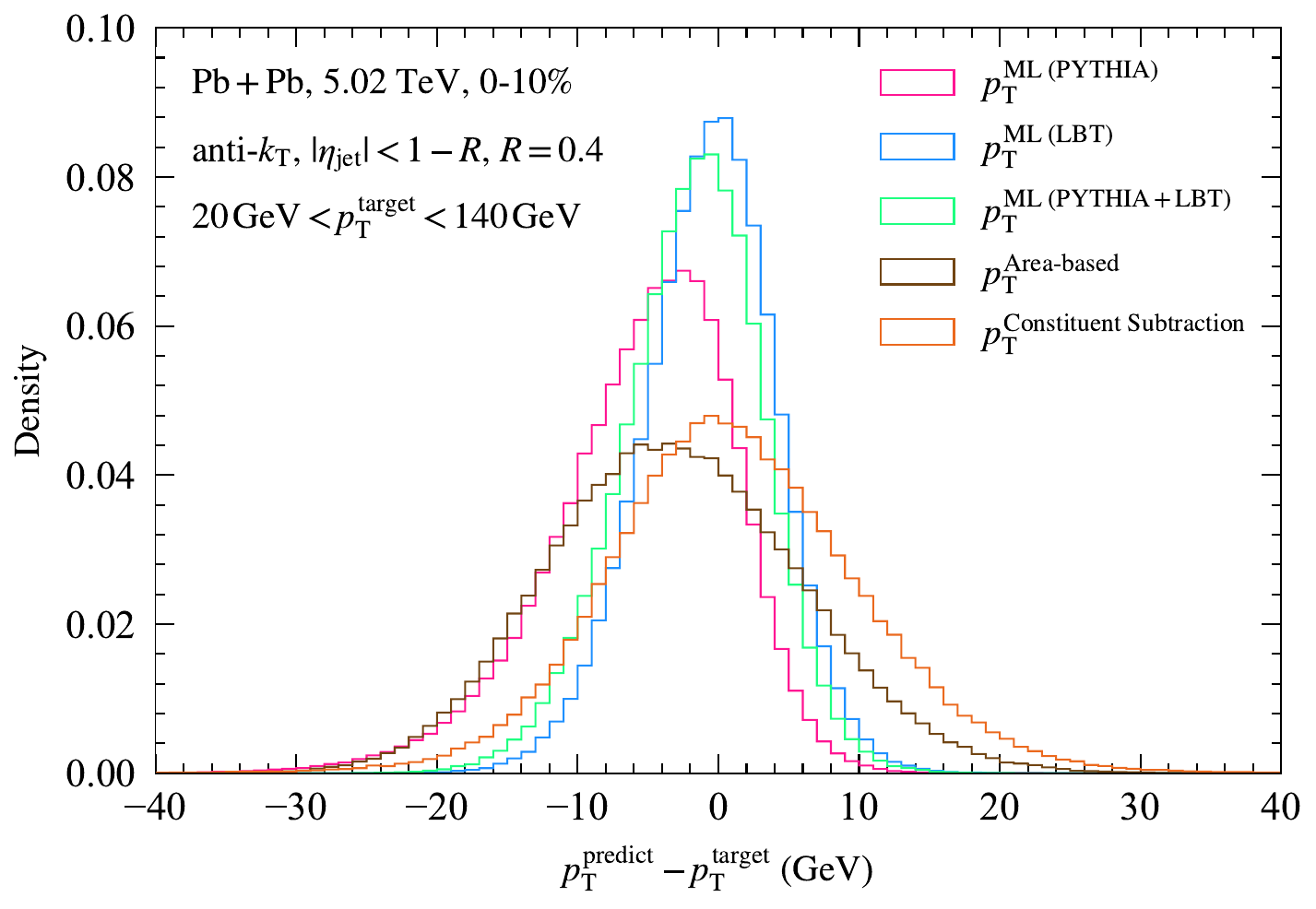}
\caption{(Color online) Residual distributions of the jet $p_\mathrm{T}$ predicted by different estimators, ML models trained on the PTYHIA, LBT, and combined PYTHIA+LBT data, and the conventional Area-based and Constituent Subtraction methods. The testing dataset is generated by the LBT model.}
\label{fig: pT_error_lbt_training} 
\end{figure}

Following Ref.~\cite{haake2019machine}, we first train the ML model on the PYTHIA data and evaluate its performance on three datasets from three different model setups: PYTHIA, LBT, and LBT without medium response particles. In the last setup, we implement the same LBT simulation as described in Sec.~\ref{sec: sim models}, except that we switch off the elastic scattering process that generates the  recoil and negative particles in our test dataset. If not otherwise specified, background particles from the thermal model are included in both training and testing data. Figure~\ref{fig: pT_error_pythia_training} presents the corresponding residual distributions of the jet $p_\mathrm{T}$, i.e., distributions of $\delta p_\mathrm{T}=p_\mathrm{T}^\mathrm{predict}-p_\mathrm{T}^{\mathrm{target}}$, with $p_\mathrm{T}^\mathrm{predict}$ the value predicted by our ML model for a jet embedded in a background, and $p_\mathrm{T}^\mathrm{target}$ its corresponding target value as defined in Eq.~(\ref{eq: pT}). Here, tests are conducted on jets with a cone size of $R=0.4$ and target $p_\mathrm{T}$ within $[20, 140]$~GeV. Note that here as well as in Figs.~\ref{fig: pT_error_lbt_training},~\ref{fig: pT_error_testing_PYTHIA} and~\ref{fig: pT_error_matching} below, results are not convoluted with the differential cross section of the initial hard collisions and therefore are not dominated by the low $p_\mathrm{T}$ jets. Such convolution will be implemented when we calculate the jet $R_\mathrm{AA}$ later. The results in Fig.~\ref{fig: pT_error_pythia_training} demonstrate the ML model trained by the PYTHIA data can accurately predict the $p_\mathrm{T}$ of the PYTHIA jets, with a mean residual close to zero and a narrow width of the distribution. However, for its prediction on the LBT jets, a notable negative bias emerges for the ML model trained by the PYTHIA data, suggesting an underestimation of the jet $p_\mathrm{T}$ due to an oversubtraction of the QGP background. This results from different inner structures between vacuum (PHTHIA) jets and medium-modified (LBT) jets. The latter contains a large number of soft medium response particles that originate from the QGP background. These particles have similar features to the background particles but should belong to medium-modified jets. Therefore, if the ML model is trained using the PYTHIA data, which do not include these medium response components, these components are very likely to be identified as background particles and subtracted from the LBT jets by mistake. To verify this cause, we further test the ML model on the LBT data with the medium response (recoil) process removed. This removal significantly improve the accuracy of the ML model prediction. The comparison shown in Fig.~\ref{fig: pT_error_pythia_training} underscores the importance of training the ML model using jet data that include jet-medium interaction effects to ensure accurate  background subtraction in heavy-ion collisions.

\begin{figure}[tbp!]
\centering
\includegraphics[width=0.48\textwidth]{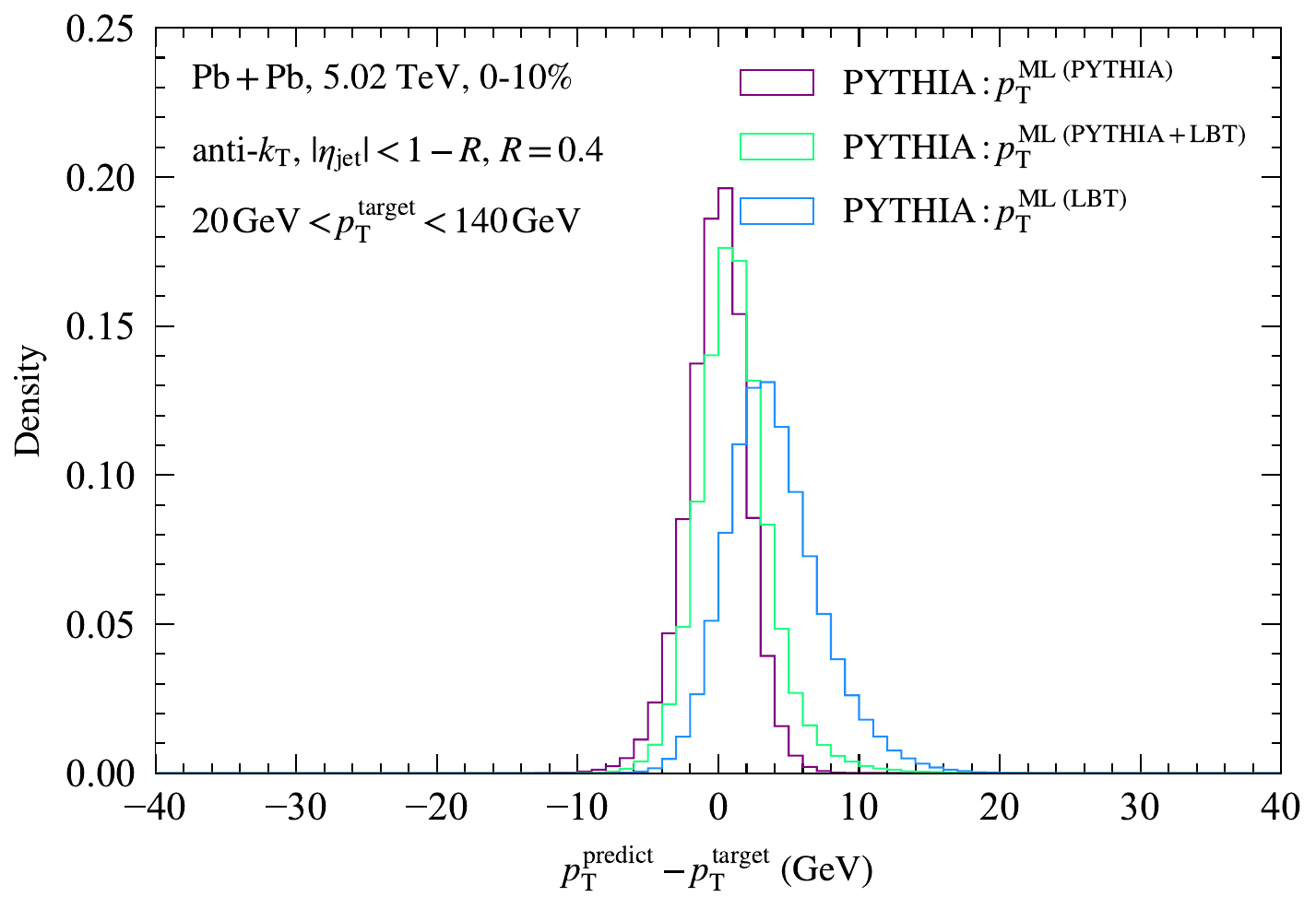}
\caption{(Color online) Residual distributions of the jet $p_\mathrm{T}$ predicted by different ML models trained on the PTYHIA, LBT, and combined PYTHIA+LBT data, respectively. The testing dataset is generated by the PYTHIA model.}
\label{fig: pT_error_testing_PYTHIA} 
\end{figure}

In Fig.~\ref{fig: pT_error_lbt_training}, we compare predictions on the jet $p_\mathrm{T}$ from different ML models and conventional background subtraction methods. Three ML models are applied, trained using the PYTHIA, LBT, and PYTHIA+LBT data, respectively, where the last dataset is the combination of the former two. All ML models are tested against the LBT data embedded in a thermal background. Again, we see a negative bias in the residual distribution of the jet $p_\mathrm{T}$ using the ML model trained on the PYTHIA data due to its incapability of recognizing the medium response particles. By training the ML model using the combined dataset from the PYTHIA and LBT simulations, this bias can be significantly reduced. The ML model trained solely on the LBT data presents the best performance, with a mean value of the distribution closest to zero and a narrowest width. The two conventional methods, the Area-based and Constituent Subtraction methods, also show oversubtractions of the background due to overestimations of the background density resulting from the contribution of recoil particles outside jet cones. The widths of the residual distributions from these two conventional methods are apparently larger than those from the LBT-trained and PYTHIA+LBT-trained ML models. Note that while $p_\mathrm{T}^\mathrm{target}$ defined in Eq.~(\ref{eq: pT}) can also be applied to evaluate the Area-based method, it cannot be applied directly to evaluate the Constituent Subtraction method in which jets are reconstructed after background subtraction. For each jet obtained via the Constituent Subtraction method, we use the matching procedure discussed in Appendix~\ref{sec: matching} to search for its nearby counterpart reconstructed with background particles, and assign the $p_\mathrm{T}^\mathrm{target}$ of the counterpart, given by Eq.~(\ref{eq: pT}), as its $p_\mathrm{T}^\mathrm{target}$. If its counterpart cannot be found, this jet does not contribute to the residual distribution in Fig.~\ref{fig: pT_error_lbt_training}. Results here highlight the advantage of the ML methods over the conventional ones on removing the QGP background for predicting the $p_\mathrm{T}$ of quenched jets in relativistic heavy-ion collisions. 

To demonstrate the robustness of the PYTHIA+LBT-trained ML model, we also present in Fig.~\ref{fig: pT_error_testing_PYTHIA} the residual distributions of the estimators ML(PYTHIA),  ML(PYTHIA+LBT), and ML(LBT) when tested on the PYTHIA data. The results show that the LBT-trained ML model tends to underestimate the background for PYTHIA jets, while PYTHIA+LBT-trained ML model provides accurate predictions with a distribution width just slightly wider than that of the PYTHIA-trained ML model. This indicates the compatibility of ML models trained on a combination of different datasets and underscores their potential for generalizability. These findings support that training ML models on combined datasets (e.g., PYTHIA and LBT or multiple jet quenching models) can significantly improve robustness and reduce biases when applied to diverse scenarios.

\begin{table}[tbp!]
  \centering
    \begin{tabular}{rrrr}
    \hline
    \hline
    \multicolumn{1}{c}{\multirow{2}[4]{*}{Estimator}} & \multicolumn{1}{c}{\multirow{2}[4]{*}{Testing Data}} & \multicolumn{2}{c}{Fitting (GeV)} \bigstrut\\
    \cline{3-4}          &       & \multicolumn{1}{c}{Mean} & \multicolumn{1}{c}{$\sigma$} \bigstrut\\
    \hline
    \multicolumn{1}{c}{ML(PYTHIA)} & \multicolumn{1}{c}{PYTHIA} & \multicolumn{1}{c}{$-0.5$} & \multicolumn{1}{c}{2.0} \bigstrut[t]\\
    \multicolumn{1}{c}{ML(PYTHIA)} & \multicolumn{1}{c}{LBT} & \multicolumn{1}{c}{$-4.8$} & \multicolumn{1}{c}{5.9} \\
    \multicolumn{1}{c}{ML(PYTHIA)} & \multicolumn{1}{c}{LBT(w/o recoil)} & \multicolumn{1}{c}{$-1.9$} & \multicolumn{1}{c}{2.8} \\
    \multicolumn{1}{c}{ML(LBT)} & \multicolumn{1}{c}{LBT} & \multicolumn{1}{c}{$-0.9$} & \multicolumn{1}{c}{4.5} \\
    \multicolumn{1}{c}{ML(PYTHIA+LBT)} & \multicolumn{1}{c}{LBT} & \multicolumn{1}{c}{$-2.0$} & \multicolumn{1}{c}{4.8} \\
    \multicolumn{1}{c}{Area-based} & \multicolumn{1}{c}{LBT} & \multicolumn{1}{c}{$-3.9$} & \multicolumn{1}{c}{9.0} \\
    \multicolumn{1}{c}{Constituent Subtract.} & \multicolumn{1}{c}{LBT} & \multicolumn{1}{c}{$0.2$} & \multicolumn{1}{c}{8.4} \bigstrut[b]\\
    \multicolumn{1}{c}{ML(LBT)} & \multicolumn{1}{c}{PYTHIA} & \multicolumn{1}{c}{$2.9$} & \multicolumn{1}{c}{3.0} \\  
    \multicolumn{1}{c}{ML(PYTHIA+LBT)} & \multicolumn{1}{c}{PYTHIA} & \multicolumn{1}{c}{$0.4$} & \multicolumn{1}{c}{2.2} \\ 
    \hline
    \hline
    \end{tabular}%
    \caption{The means and standard deviations of the residual distributions of the jet $p_\mathrm{T}$ from different estimators evaluated on different testing datasets.}
  \label{tab: mean}%
\end{table}%

For a quantitative comparison between these methods, Tab.~\ref{tab: mean} lists the means and standard deviations ($\sigma$) of the residual distributions presented in Figs.~\ref{fig: pT_error_pythia_training}--\ref{fig: pT_error_testing_PYTHIA} for different estimators evaluated on different testing datasets.  One may notice that the performance of the PYTHIA-trained ML model on the PYTHIA jets is much better than that of the LBT-trained ML model on the LBT jets. This can be understood with the more complicated structures of medium-modified jets from LBT than vacuum jets from PYTHIA. The LBT jets experience quenching inside the QGP medium to quite different degrees due to different traversing lengths, while the PYTHIA jets can be viewed as the no-quenching limit of the LBT jets. As discussed for Fig.~\ref{fig: pT_error_pythia_training}, medium-modified jets inevitably include recoil particles that originate from the thermal background, but are considered part of jets. This introduces additional difficulty for the ML methods to separate background and jets. The comparison in Tab.~\ref{tab: mean} demonstrates the superior performance of the ML model trained on the LBT data to other methods when tested on LBT data. Although the ML model trained on the combined PYTHIA+LBT data loses certain accuracy compared to that trained solely on the LBT data, the former could be more flexible than the latter in application to diverse environments, e.g., different centralities, of heavy-ion collisions. 

We note that in our current setup, patterns of medium response can enter via various substructure observables of jets, such as the jet mass, angularity and momentum dispersion. Recently, the energy-energy correlator (EEC)~\cite{Yang:2023dwc,Xing:2024yrb,Barata:2024ukm} and multi-point energy correlators (ENC)~\cite{Bossi:2024qho,Barata:2025fzd} have also been shown sensitive to medium response. Methods of cancelling effects of medium response on energy correlators have been explored~\cite{Zhao:2025ogc} but have not been confirmed by realistic calculations yet. Incorporating these novel observables as input would allow our ML models to grasp more comprehensive features of medium modified jets and therefore isolate them from the QGP background with higher precision.

\section{Nuclear Modification Factor of Jets} 
\label{sec: RAA}

\begin{figure}[tbp!]
\centering
\includegraphics[width=0.48\textwidth]{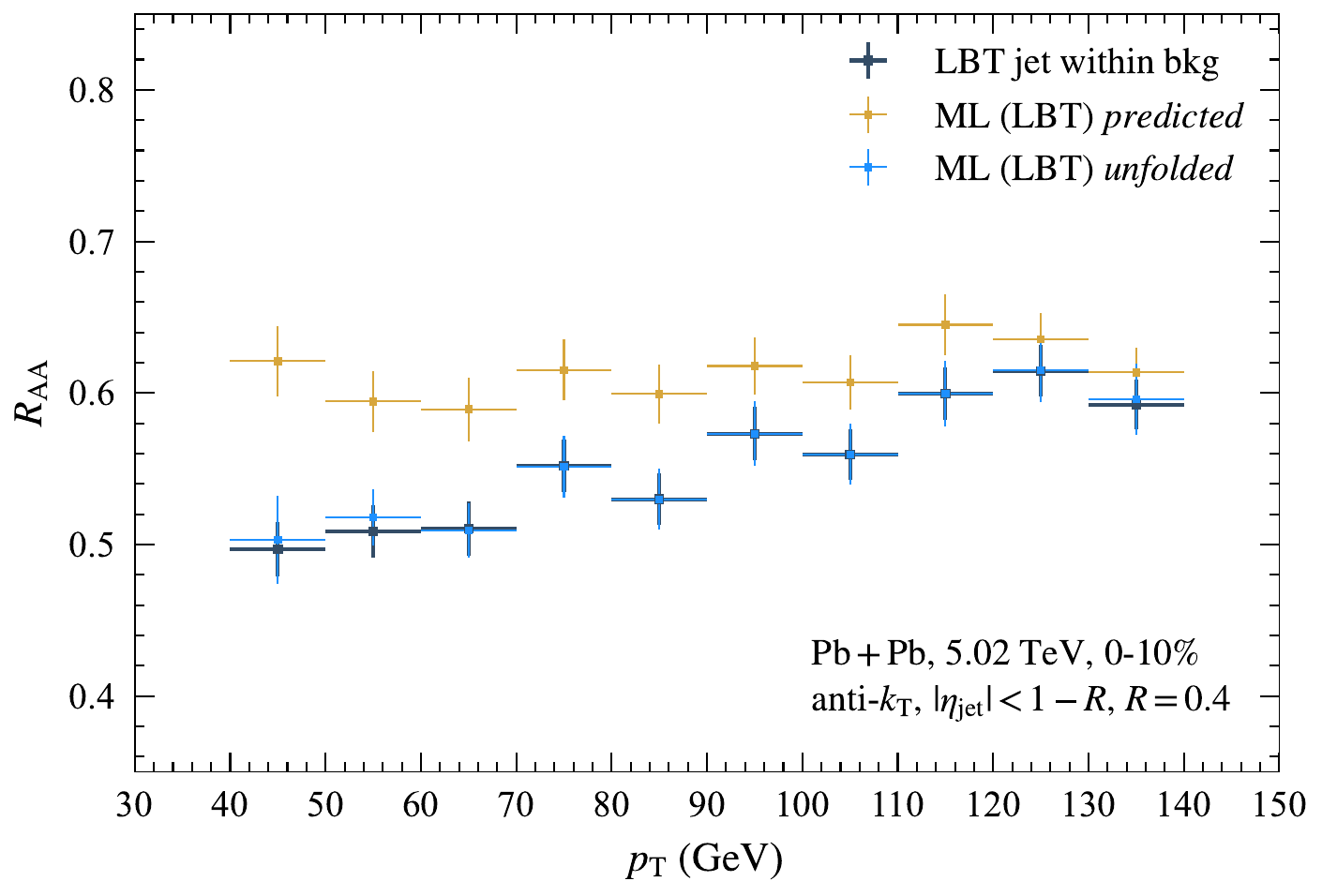}
\caption{(Color online) The nuclear modification factor of jets obtained from the ML model trained on the LBT dataset, compared between with and without the unfolding correction, and a baseline of the LBT jets reconstructed inside a QGP background (LBT jet within bkg).}
\label{fig: Raa_0} 
\end{figure}

The nuclear modification factor ($R_\mathrm{AA}$), defined as the ratio of the jet spectrum in A+A collisions, scaled by the average number of nucleon-nucleon collisions, to that in $p+p$ collisions, measures the amount of jet quenching in high-energy heavy-ion collisions. Based on the jet $p_\mathrm{T}$ predicted by the ML models above, we further study the performance of these models on evaluating the jet $R_\mathrm{AA}$. Given the steeply falling $p_\mathrm{T}$ spectra of jets, an unfolding procedure is essential to reveal the jet quenching effect accurately. In this work, we employ the iterative Bayesian unfolding algorithm~\cite{DAgostini:2010hil} with four iterations, which provides a balance between bias reduction and statistical fluctuations for our kinematic range. For each ML model, the response matrix of unfolding is constructed between the true $p_\mathrm{T}$, or $p_\mathrm{T}^\mathrm{target}$ from Eq.~(\ref{eq: pT}), of jets and their $p_\mathrm{T}$ predicted by the ML model.

In Fig.~\ref{fig: Raa_0}, we present the jet $R_\mathrm{AA}$ predicted by the LBT-trained ML model. The jet spectrum in $p+p$ collisions is obtained directly from the PYTHIA simulation, and their spectrum in Pb+Pb collisions is calculated using the ML model predicted $p_\mathrm{T}$ of LBT jets embedded in the thermal background. Here, the ML model results are also compared to a baseline (``LBT jet within background") that is obtained using the $p_\mathrm{T}^\mathrm{target}$ of the LBT jets reconstructed inside the background. One can observe that while the LBT-trained ML model predicts the jet $p_\mathrm{T}$ well, it still overestimates the jet $R_\mathrm{AA}$ if its predicted $p_\mathrm{T}$ values are directly used to calculate the jet $R_\mathrm{AA}$. This results from the convolution of its prediction residual with the steeply falling $p_\mathrm{T}$ spectrum of jets. When the unfolding procedure is applied, this overestimation can be effectively reduced.

\begin{figure}[tbp!]
\centering
\includegraphics[width=0.48\textwidth]{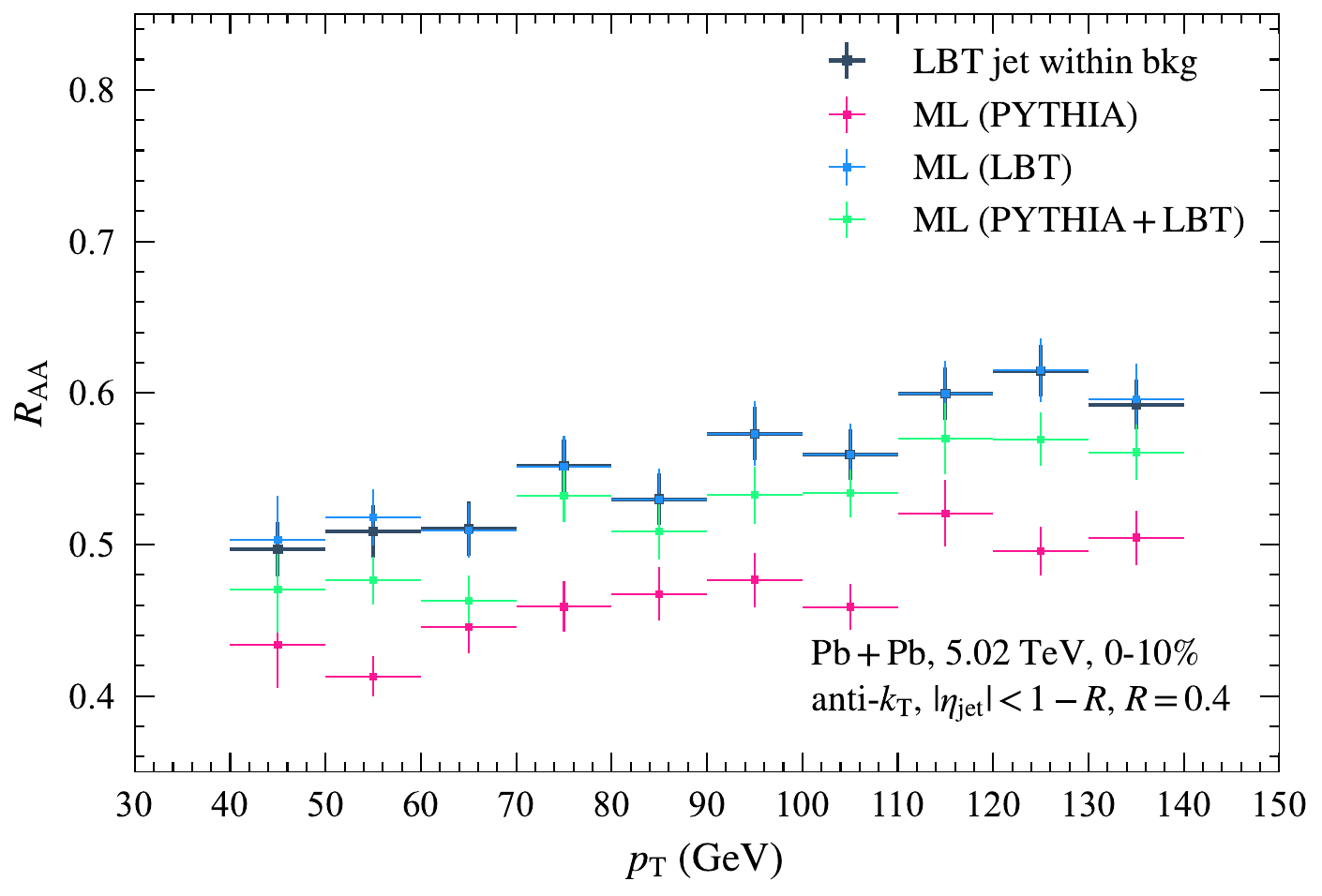}
\caption{(Color online) The nuclear modification factor of jets obtained from ML models trained on the PYTHIA, LBT, and combined PYTHIA+LBT datasets with the unfolding correction, compared to a baseline of the LBT jets reconstructed in the presence of a QGP background (LBT jet within bkg).}
\label{fig: Raa_1} 
\end{figure}

In Fig.~\ref{fig: Raa_1}, we further compare the unfolded jet $R_\mathrm{AA}$ between different ML models, trained on the PYTHIA, LBT, and combined PYTHIA+LBT data, respectively. To compare the accuracy of these ML models, the baseline of ``LBT jet within background" is also presented. Although all ML models aim at approximating the results of the LBT jets within the QGP background, the PYTHIA-trained ML model significantly underestimates the $R_\mathrm{AA}$, which primarily stems from its oversubtraction of the background as discussed earlier and persists even after unfolding. In contrast, the PYTHIA+LBT-trained ML model provides an evaluation much closer to the baseline, and the LBT-trained ML model achieves an excellent agreement. While the PYTHIA+LBT-trained and LBT-trained ML models demonstrate comparable performance in reconstructing the $p_\mathrm{T}$ of the LBT jets, their unfolded $R_\mathrm{AA}$'s exhibit a systematic discrepancy, which stems from the distinctions in their respective response matrices. Specifically, the PYTHIA+LBT-trained ML model employs a hybrid response matrix combining jet spectra from both PYTHIA and LBT simulations, introducing a domain mismatch between training and application; while the LBT-trained ML model employs a dedicated response matrix derived exclusively from LBT simulations, maintaining domain consistency between training and application. In Ref.~\cite{acharya2024measurement}, ML models trained separately on PYTHIA and its variants that mimic quenching effects are utilized to extract the jet $R_\mathrm{AA}$ with associated uncertainties from the experimental data. Based on our findings, we recommend either excluding PYTHIA jets from separate training datasets or at most combining them with realistic quenched jets samples to mitigate bias. Training ML models on jet data that capture jet quenching effects is essential for enhancing the reliability of these models in predicting jet observables. We have also verified that the jet $R_\mathrm{AA}$ obtained using our LBT-trained and PYTHIA+LBT-trained ML models excel that obtained using the Area-Based and Constituent Subtracting methods in subtracting the QGP background.

In principle, when the QGP background and its fluctuation are strong, there could be mismatch between jets reconstructed with and without the background particles. This includes clustering different jet partons into a jet and clustering background particles into fake jets, and therefore introduces additional uncertainties to jet observables. In Appendixes~\ref{sec: matching} and~\ref{sec: LBT-only jets}, we will show that this mismatch has minor impact on the jet $R_\mathrm{AA}$, and thus the ML models together with the unfolding correction we develop in this study are sufficient for studying the jet quenching effect in the presence of the QGP background.

\section{Conclusions} 
\label{sec: Conclusions}

We have applied the machine learning techniques on reconstructing jet momentum within the QGP background in relativistic heavy-ion collisions. By training the DNN models on simulated data from both PYTHIA and LBT models for comparative analysis, we have demonstrated that the ML-based reconstruction, when properly taking into account the jet quenching effects, achieves high accuracy in jet momentum reconstruction within the QGP background. In particular, while the model trained on the PYTHIA data accurately predicts the $p_\mathrm{T}$ of vacuum (PYTHIA) jets, it exhibits significant bias when being applied to medium-modified (LBT) jets, primarily due to an oversubtraction of the medium response particles from jet-medium interactions, which resemble the background particles but belong to jets. However, by training on the quenched jet data generated by the LBT model, the DNN model effectively reduces this prediction bias and performs a more accurate background subtraction compared to the conventional Area-based and Constituent Subtraction methods. The DNN model trained on a combined PYTHIA+LBT dataset appears equally effective, and is generalizable to more diverse environments.

This improvement is also highlighted in our analyses of the nuclear modification factor of jets, where the LBT-trained ML model and the PYTHIA+LBT-trained ML model excel in approximating the baseline of the reconstructed LBT jets within the QGP background, compared to the PYTHIA-trained ML model, across a wide $p_\mathrm{T}$ range. The convolution between the residual distribution of the ML model predicted jet $p_\mathrm{T}$ and the steeply falling $p_\mathrm{T}$ spectra of jets can lead to an overestimation of the jet $R_\mathrm{AA}$, which can be effectively corrected using an unfolding procedure. However, the superior performance of the ML models trained using the quenched jet data to that trained using the vacuum jet data still holds even after the unfolding correction is applied.

This work not only demonstrates that machine learning provides a powerful approach for jet momentum reconstruction within a complicated QGP background, but also helps interpret the physics origins of its successful performance. Our results underscore the importance of training models on quenching-aware datasets and combination of multiple datasets to ensure accurate background subtraction in relativistic heavy-ion collisions, and extend a reliable prediction on the jet $R_\mathrm{AA}$ down to $p_\mathrm{T}$ around 40~GeV, which can be hardly achieved by conventional methods. 

While the present results are based on training with the LBT model and inevitably retain model dependence, they remain qualitatively robust and provide valuable physics insights. Future efforts will aim to reduce this dependence by training ML models on combined datasets from multiple jet quenching models and vacuum jet samples with varied fragmentation patterns, together with realistic heavy-ion backgrounds including collective flows. The resulting models will then be validated against datasets from unseen quenching models to directly assess their generalizability. In parallel, a promising future direction is to integrate full jet information and prior physical knowledge into the machine-learning framework. At the background energy subtraction level, beyond the point cloud representation of jet constituents, theoretically motivated observables such as fragmentation functions, jet shapes, or energy correlators could be incorporated as additional inputs to improve the jet $p_\mathrm{T}$ reconstruction. At the background particle removal level, jet substructure observables (e.g., groomed splitting angles, momentum sharing between subjets and energy correlators) may be embedded directly into the training objectives alongside the jet momentum. Such a physics-informed paradigm would steer the model to learn fundamental jet-medium dynamics rather than superficial correlations, thereby complementing the multi-model training strategy above and enhancing both interpretability and robustness of ML-based jet reconstruction. We view this combination of data-driven and theory-driven strategies as a natural and promising pathway for advancing ML-based background particle removal and jet observables reconstruction toward robust applications in heavy-ion collisions, enabling a deeper understanding of the QGP properties and advancing the field of high-energy nuclear physics.

\section*{Acknowledgments}
We are grateful to helpful discussions with Yang He, Maowu Nie and Long-Gang Pang. This work is supported in part by the National Natural Science Foundation of China (NSFC) under Grant Nos. 12175122, 2021-867 (R.L., S.C.), and in part by the Taishan Scholars Program under Grant No. tsqnz20221162, and Shandong Excellent Young Scientists Fund Program (Overseas) under Grant No. 2023HWYQ-106 (Y.D.).

\appendix

\section{Jet matching}
\label{sec: matching}

As discussed in Sec.~\ref{sec:const}, a strong medium background could introduce clustering bias. In other words, jet partons which should be clustered into one jet in the absence of the QGP background may no longer belong to the same jet when the background particles also participate in jet clustering. Therefore, even the Area-based method or the ML method can cleanly remove the background components within a jet reconstructed inside a background, it cannot accurately reproduce properties of jets that have never been contaminated by the background. For convenience, we will call jets reconstructed in the absence of the QGP background ``real" jets or ``LBT-only" jets. Meanwhile, background fluctuations may lead to ``fake" jets, which have no connection to real jets originating from the initial nucleon-nucleon hard scatterings. All the three background subtraction methods above can suffer from these fake jets especially for low $p_\mathrm{T}$ jets. 

To estimate the error introduced by such jet mismatching and further improve the ML performance on predicting the  $p_\mathrm{T}$ of ``real" jets, we employ the following jet matching procedure. For each jet reconstructed inside a background $j_\text{in bkg}$, we search for its counterpart $j_\text{real}$ from its corresponding PYTHIA or LBT event in the absence of the background. We require the distance between $j_\text{in bkg}$ and $j_\text{real}$ satisfy $\Delta R = \sqrt{\Delta\eta^2 + \Delta\phi^2} < 0.4$. If multiple jets from the PYTHIA or LBT event satisfy this condition, the one with the smallest $\Delta R$ is identified as $j_\text{real}$ for $j_\text{in bkg}$. On the other hand, if no $j_\text{real}$ can be found, $j_\text{in bkg}$ is identified as a fake jet. The ML method discussed in Sec.~\ref{sec: ML method} can then be updated by replacing its target value from Eq.~(\ref{eq: pT}) to the $p_\mathrm{T}$ of $j_\text{real}$. Effects of this update will be discussed in Appendix~\ref{sec: LBT-only jets}. 

The same matching procedure is also applied in Sec.~\ref{sec:momentum} for evaluating the performance of the Constituent Subtraction method in reconstructing the jet momentum. Considering that jets are reconstructed after background subtraction in this method, the $p_\mathrm{T}^\mathrm{target}$ of jets cannot be directly obtained using Eq.~(\ref{eq: pT}). Therefore, we match each jet reconstructed using the Constituent Subtraction method to a jet reconstructed together with the background particles and use the $p_\mathrm{T}^\mathrm{target}$ of the latter as the $p_\mathrm{T}^\mathrm{target}$ of the former.

\section{Impact of mismatch on the jet $R_\mathrm{AA}$}
\label{sec: LBT-only jets}

\begin{figure}[tbp!]
\centering
\includegraphics[width=0.48\textwidth]{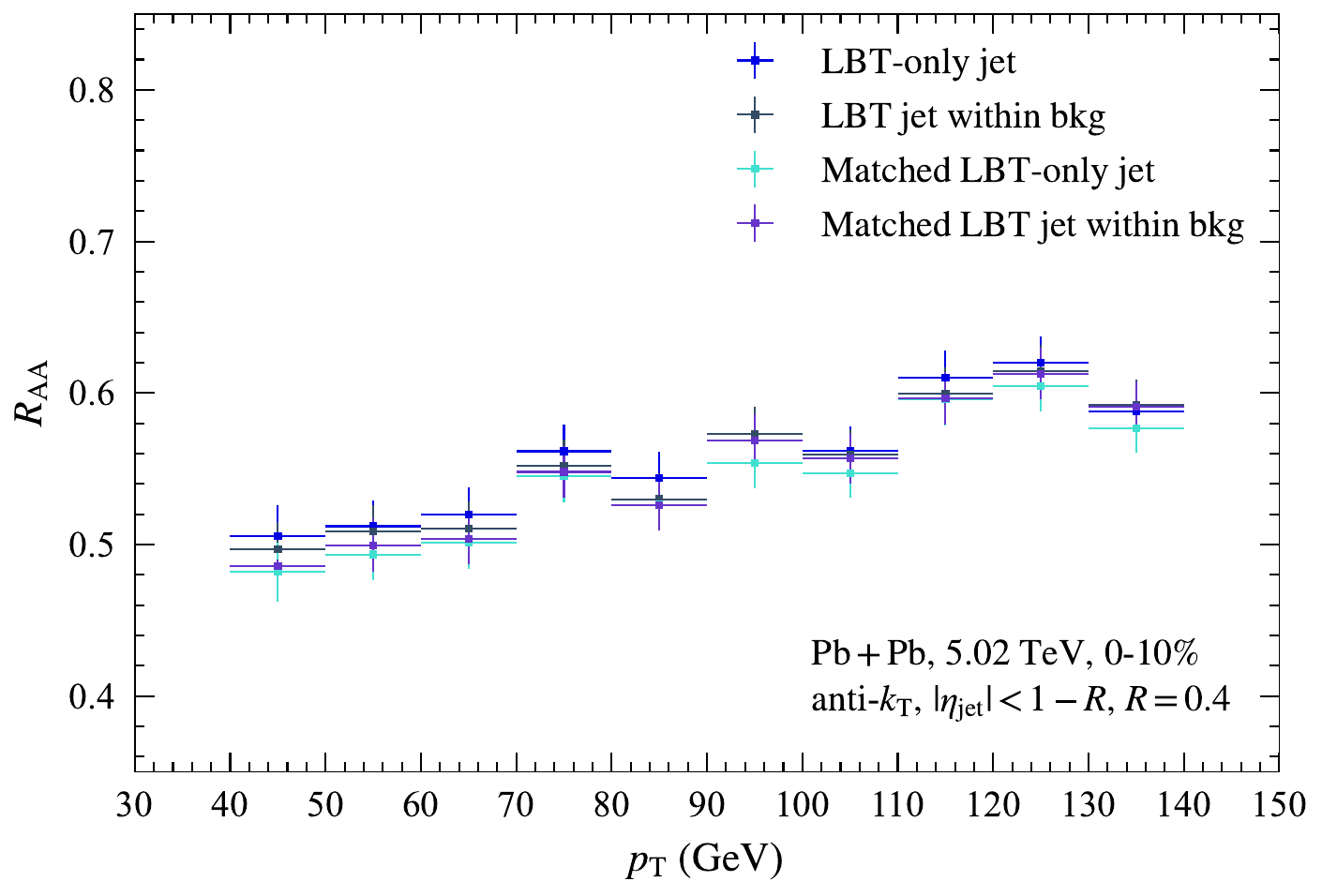}
\caption{(Color online) The nuclear modification factors of jets obtained from the LBT jets reconstructed with (LBT jet within bkg) and without (LBT-only jet) the background particles as two baselines and their successfully matched counterparts, i.e., Matched LBT jet within bkg and Matched LBT-only jet.}
\label{fig: Raa_3} 
\end{figure}

Following discussions in Appendix~\ref{sec: matching}, ``LBT jet within bkg" presented in Figs.~\ref{fig: Raa_0} and~\ref{fig: Raa_1} may not be the exact baseline of jets from theoretical calculations in literature. These jets within background are first reconstructed together with the background particles, and then their $p_\mathrm{T}$ are calculated using their jet parton components. However, the presence of a strong medium background could bias the clustering procedure in the first step, and therefore removing the background contribution in the second step cannot guarantee that $j_\text{in bkg}$ returns to $j_\text{real}$. In Fig.~\ref{fig: Raa_3}, we add another baseline of the LBT jets, which are reconstructed by using jet partons solely from the LBT model output, without being placed inside a medium background. This baseline of $R_\mathrm{AA}$, labeled as ``LBT-only jet", is expected to reflect the medium modification of jets in theoretical calculations. We see a slight discrepancy between ``LBT-only jet" and ``LBT jet within bkg", indicating that reconstructed jets which are mismatched to their real counterparts only affect the accuracy of jet observables to a tolerable degree.

\begin{figure}[tbp!]
\centering
\includegraphics[width=0.48\textwidth]{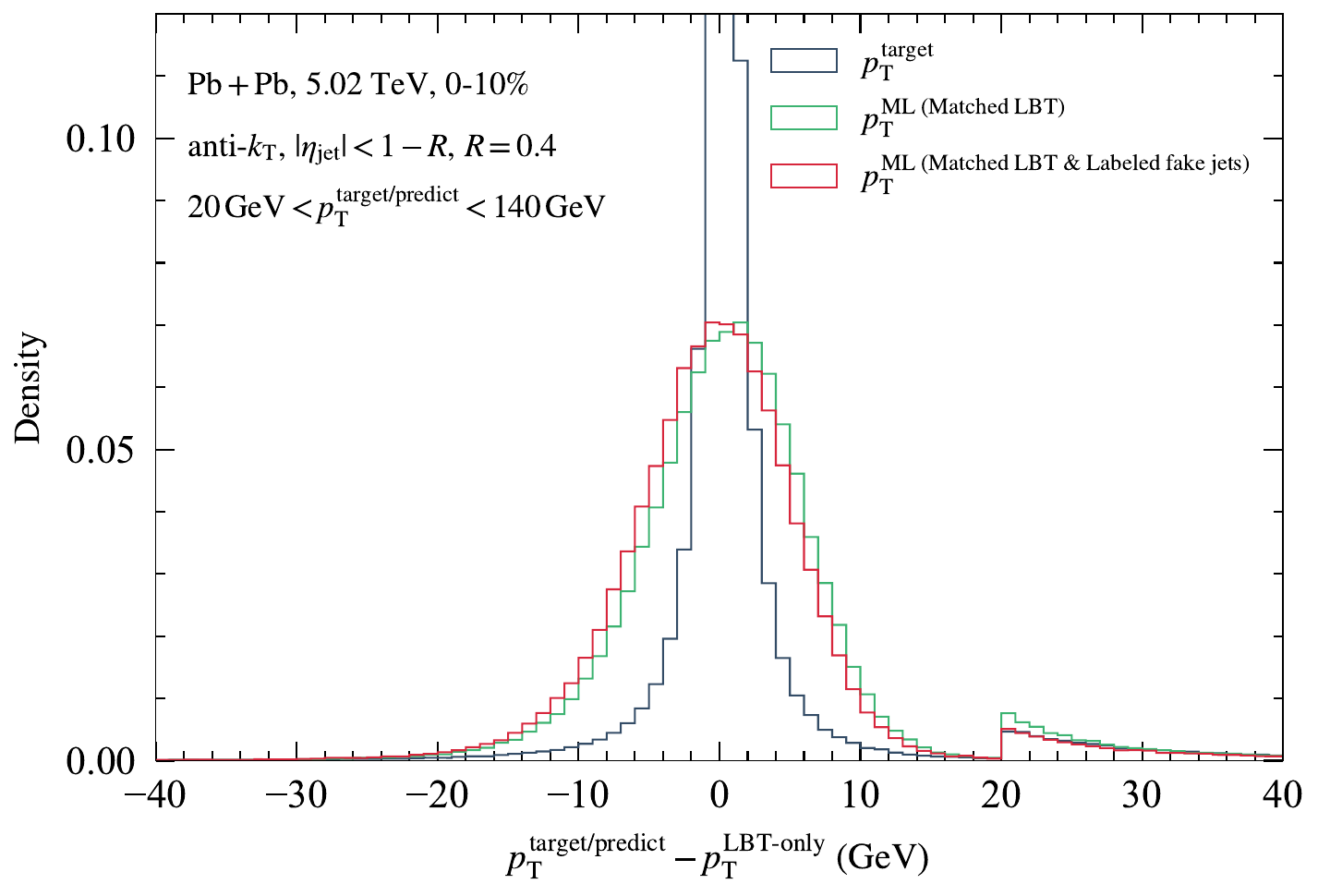}
\caption{(Color online) Residual distributions of the jet $p_\mathrm{T}$ predicted by different ML models trained on the Matched LBT and Matched LBT $\&$ Labeled fake jets. The comparisons are made with the $p_\mathrm{T}$ of the matched LBT-only jets. The target $p_\mathrm{T}$ value obtained using Eq.~(\ref{eq: pT}) is also shown for comparison. The abrupt changes at 20 GeV in all these curves indicate the involvement of fake jets since we label the $p_\mathrm{T}$ of their LBT-only counterparts as zero.}
\label{fig: pT_error_matching} 
\end{figure}

To tackle this discrepancy in estimating the jet $R_\mathrm{AA}$ in detail, we apply the matching procedure introduced in Appendix~\ref{sec: matching} to pair the LBT jets reconstructed inside the QGP background to those reconstructed without the QGP background. After this matching, we present the $R_\mathrm{AA}$ calculated using the $p_\mathrm{T}$ of matched LBT jets within background (``Matched LBT jet within bkg") and using the $p_\mathrm{T}$ of matched LBT-only jets (``Matched LBT-only jet") in Fig.~\ref{fig: Raa_3}. We see that ``Matched LBT jet within bkg" and ``Matched LBT-only jet" can reproduce the ``LBT jet within bkg" and ``LBT-only jet" baselines well, respectively. The slight underestimation of the $R_\mathrm{AA}$ baselines for both cases indicates that most LBT-only jets can be paired from reconstructed jets inside the QGP background via this matching procedure, while a few of them fail the matching due to large deflection of their jet axes by including too many soft background particles during jet reconstruction. Meanwhile, a few of these reconstructed jets inside the QGP background also fail the matching and are identified as the ``fake" jets. 

The discrepancy of the jet $R_\mathrm{AA}$ between baselines of LBT-only jets and LBT jets within the background could also stem from the $p_\mathrm{T}$ difference between the matched jets. We present the residual distribution of the $p_\mathrm{T}$ of these matched jets labeled by $p_\text{T}^{\mathrm{target}}-p_\text{T}^{\mathrm{LBT-only}}$ in Fig.~\ref{fig: pT_error_matching}. The result shows that the $p_\mathrm{T}$ of matched LBT-only jets tend to be slightly higher than that of their counterparts reconstructed inside the QGP background. 

\begin{figure}[tbp!]
\centering
\includegraphics[width=0.48\textwidth]{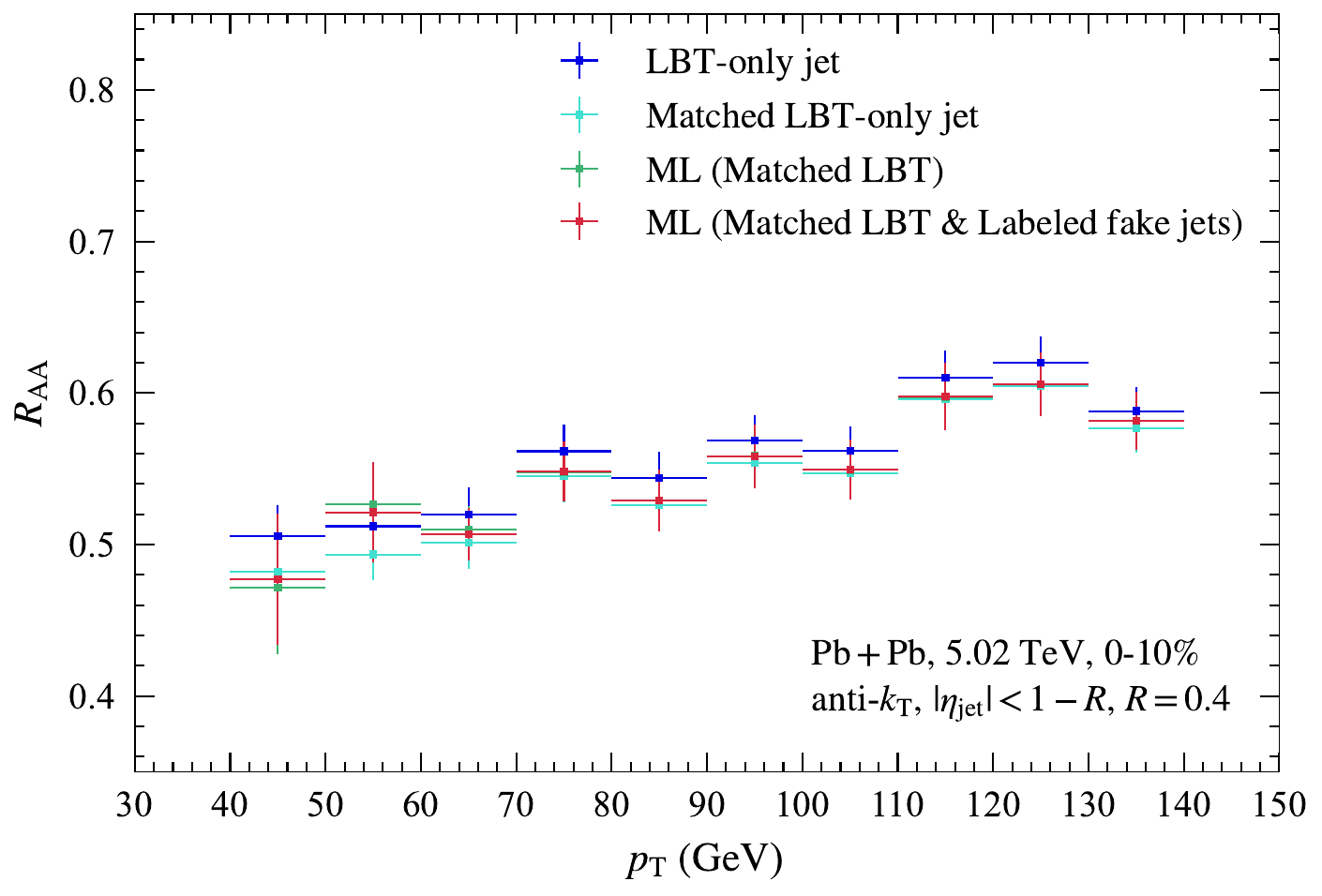}
\caption{(Color online) The nuclear modification factors of jets obtained from ML models trained on the matched LBT jet data, with and without labeling the fake jets, compared to the baseline of the LBT jets reconstructed without the background (LBT-only jet) and a direct calculation using the matched LBT jets (Matched LBT-only jet).}
\label{fig: Raa_4} 
\end{figure}

With the matching procedure, we use the $p_\mathrm{T}$ of these matched LBT-only jets, instead of Eq.~(\ref{eq: pT}), as the target values to re-train our ML models in two different scenarios. For the first model, we only include reconstructed jets which can be successfully matched to LBT-only jets in the training dataset, while excluding reconstructed jets whose LBT-only counterparts cannot be found. The residual distribution of the predicted $p_\mathrm{T}$ by this newly-trained ML model is also presented in Fig.~\ref{fig: pT_error_matching}, labeled as ``ML (Matched LBT)". The corresponding $R_\mathrm{AA}$ result is presented in Fig.~\ref{fig: Raa_4}, which agrees well with that of the LBT-only jet baseline. For the second model, besides labeling the $p_\mathrm{T}$ of reconstructed jets inside the background as the $p_\mathrm{T}$ of their matched LBT-only jets, as we did for the first model,  we label the $p_\mathrm{T}$ of fake jets as zero during the model training. The residual distribution of this ML model, labeled as ``Matched LBT \& Labeled fake jets", is slightly shifted leftwards compared with that of ML (Matched LBT) as shown in Fig.~\ref{fig: pT_error_matching}. The corresponding $R_\mathrm{AA}$ obtained using this ML model also agrees with the LBT-only baseline. The minor difference between labeling and not labeling fake jets in training ML models implies the negligible impact of fake jets on the jet $R_\mathrm{AA}$. Although after introducing the matching procedure, the jet $R_\mathrm{AA}$'s predicted by the two ML models here appear better than the one presented in Fig.~\ref{fig: Raa_0} prior to the unfolding correction, we have verified that after the unfolding correction, no apparent difference can be seen in their performance on predicting the jet $R_\mathrm{AA}$. Effects of this jet mismatch on other jet observables will be left for our future investigation.

\newpage

\bibliographystyle{apsrev4-1}
\bibliography{duyl}

\end{document}